\documentclass[sigconf,nonacm,9pt]{acmart}
\usepackage{graphicx}
\usepackage{booktabs}
\usepackage{multirow} 
\usepackage{cuted}     % 终极魔法：用于打破双栏
\usepackage{capt-of}   % 终极魔法：用于非浮动体标题

\setcopyright{acmlicensed}
\usepackage{enumitem}
\renewcommand\footnotetextcopyrightpermission[1]{}

\title{Screen, Cache, and Match: A Training-Free Causality-Consistent Reference Frame Framework for Human Animation}

\author{
Jianan~Wang$^{1}$,
Nailei~Hei$^{1}$,
Li~He$^{1}$,
Huanzhen~Wang$^{1}$,
Aoxing~Li$^{1}$,
Yingkai~Zhao$^{1}$,
Yuxuan~Lin$^{2}$,
Haofen~Wang$^{3}$,
Chunyang~Wang$^{4}$,
Yan~Wang$^{4}$,
Wenqiang~Zhang$^{1,2}$\\  % 换行分隔作者与单位
\small  % 单位字体缩小（可选，提升排版美观度）
$^{1}$College of Intelligent Robotics and Advanced Manufacturing, Fudan University, Shanghai, China\\
$^{2}$Shanghai Key Lab of Intelligent Information Processing, College of Computer Science and Artificial Intelligence, Fudan University, Shanghai, China\\
$^{3}$College of Design and Innovation, Tongji University, Shanghai, China\\
$^{4}$School of Data Science and Engineering, East China Normal University, Shanghai, China\\}
\renewcommand{\authors}{Jianan Wang et al.}
\renewcommand{\shortauthors}{Wang et al.}

\begin{document}
\begin{teaserfigure}
  \vspace{-13pt}
  \centering
 \includegraphics[width=\textwidth,height=0.575\textheight,keepaspectratio]{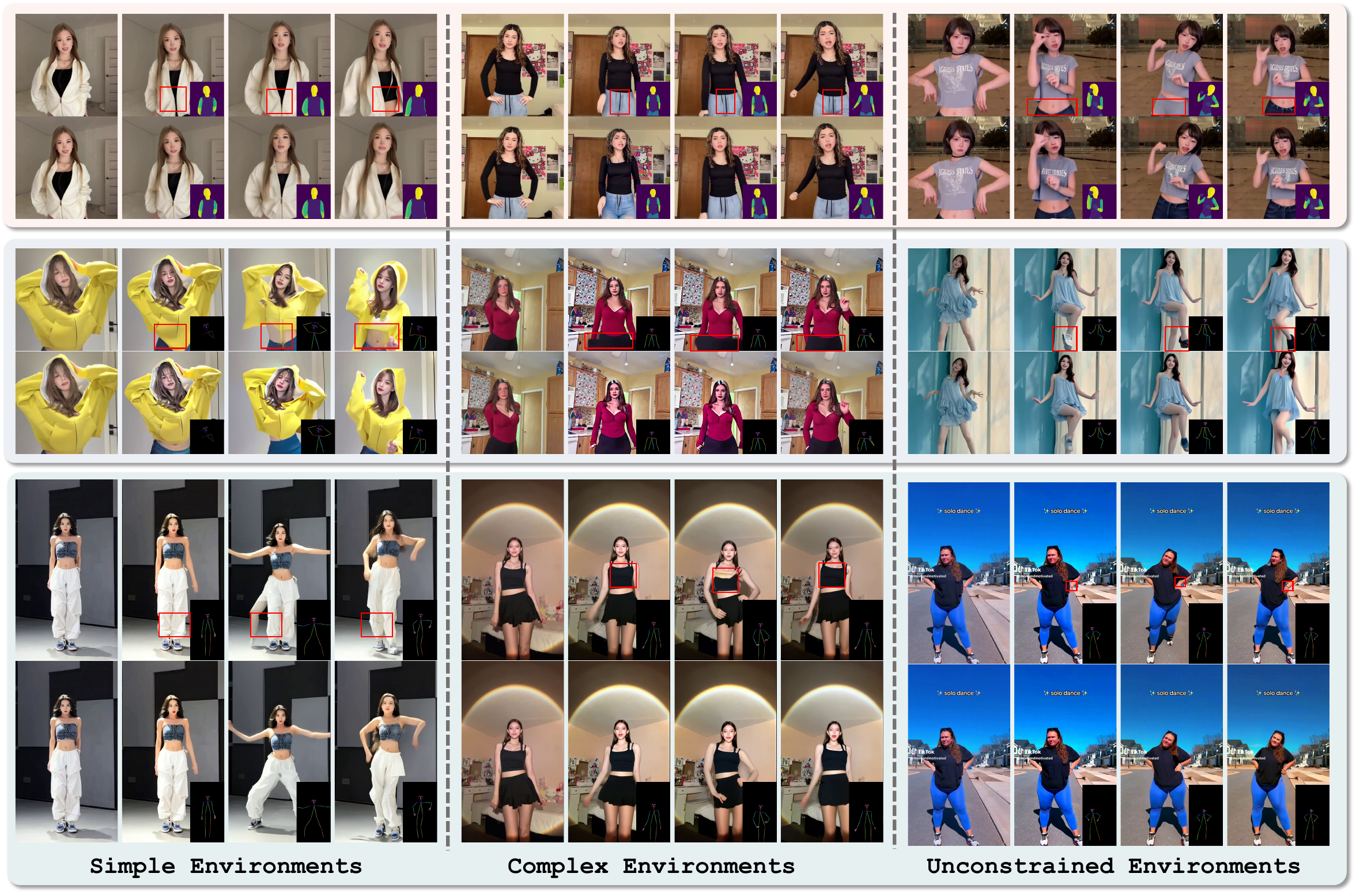}
  \caption{Visual comparison of original animations (top) and FrameCache-enhanced results (bottom) across three baselines.}
  \label{fig:fig1}
\end{teaserfigure}
\begin{abstract}
Human animation aims to generate temporally coherent and visually consistent videos over long sequences, yet modeling long-range dependencies while preserving frame quality remains challenging. Inspired by the human ability to leverage past observations for interpreting ongoing actions, we propose FrameCache, a training-free, causality-consistent reference frame framework. FrameCache explicitly converts historical generation results into causal guidance through two complementary mechanisms. First, at the reference level, a novel Screen-Cache-Match (SCM) strategy constructs a dynamic, high-quality reference memory, ensuring motion-consistent appearance guidance to reduce identity drift. Second, at the generative level, a Trajectory-Aware Autoregressive Generation (TAAG) mechanism aligns denoising trajectories across adjacent video chunks. This is achieved through an overlap-aware latent propagation and a dual-domain fusion strategy that seamlessly blends low-frequency structural layouts with high-frequency textural details. Extensive experiments on standard benchmarks demonstrate that FrameCache consistently improves temporal coherence and visual stability while integrating seamlessly with diverse diffusion baselines. Code will be made publicly available.
\end{abstract}

\maketitle
\vspace{-4pt}
\section{Introduction}
Character animation is a central direction in generative artificial intelligence, aiming to synthesize realistic and high-quality video sequences from reference images and target pose sequences. Beyond visual fidelity, these techniques have significant potential in robotics and human–robot interaction (HRI), supporting applications such as teleoperation, embodied avatars, collaborative robot learning, and digital twins~\cite{2017Realtime}. Recent advances are largely driven by diffusion-based frameworks~\cite{ho2020denoising,zhu2024champ,wang2024disco,hu2023animateanyone,hu2025animate,karras2023dreampose,tan2025animate,tu2025flashportrait,zhang2025flexiact,hong2025free}, which leverage latent diffusion models with conditional guidance to improve appearance consistency, motion stability, and generalization across diverse characters. These methods have demonstrated remarkable progress in short-sequence animation generation, setting the foundation for long-horizon tasks in real-world interactive scenarios.

Despite these advances, long-sequence character animation remains an unsolved challenge. Existing methods often rely on a single reference image combined with pose or motion signals to generate future frames~\cite{hu2023animateanyone,zhang2025mimicmotion,xu2024magicanimate,chang2023magicpose,wang2025unianimate,tu2025stableanimator,xue2025human,xu2025hypermotion,ding2025mtvcrafter,shi2025one,he2025posegen}. While effective for short sequences or simple motions, these approaches struggle when faced with complex motion patterns, occlusions, or large pose variations. Such scenarios often lead to detail blurring, temporal inconsistency, and semantic ambiguity, which accumulate over long sequences and degrade both visual fidelity and identity preservation. For example, fine-grained details such as clothing accessories, facial features, or background elements may inconsistently appear or disappear across frames. In addition, many existing methods demonstrate limited generalization: they perform well on specific characters or motion types seen during training but fail to maintain temporal consistency across unseen motions or diverse character appearances. This shortcoming underscores a fundamental gap: current generative frameworks lack mechanisms to explicitly model long-term temporal dependencies while preserving fine-grained visual details.

As illustrated in Fig.~\ref{fig:fig1}, these challenges are exacerbated under large motion variations. Consider a scenario where a person performs large-scale movements, causing the hem of a half-zipped jacket to undergo drastic motion: relying solely on the original reference frame typically leads to unrealistic temporal variations in clothing appearance across frames. Similarly, when previously occluded or unseen regions (e.g., the waistband of pants) become visible due to pose changes, existing models may inconsistently hallucinate these areas, resulting in visually jarring temporal discontinuities. These failures reflect the absence of effective strategies to jointly reason over temporal correlations and spatial details, leading to errors that compound over time and undermine identity and appearance consistency. This gap motivates a central question: how can generative systems dynamically leverage previously generated frames or reference images to guide long-horizon animation while maintaining temporal and visual coherence? Addressing this question is crucial for both aesthetic quality and practical applications where even minor temporal inconsistencies can accumulate into noticeable artifacts and affect task reliability.
\begin{figure}[htbp]
  \centering
  \includegraphics[width=0.95\linewidth]{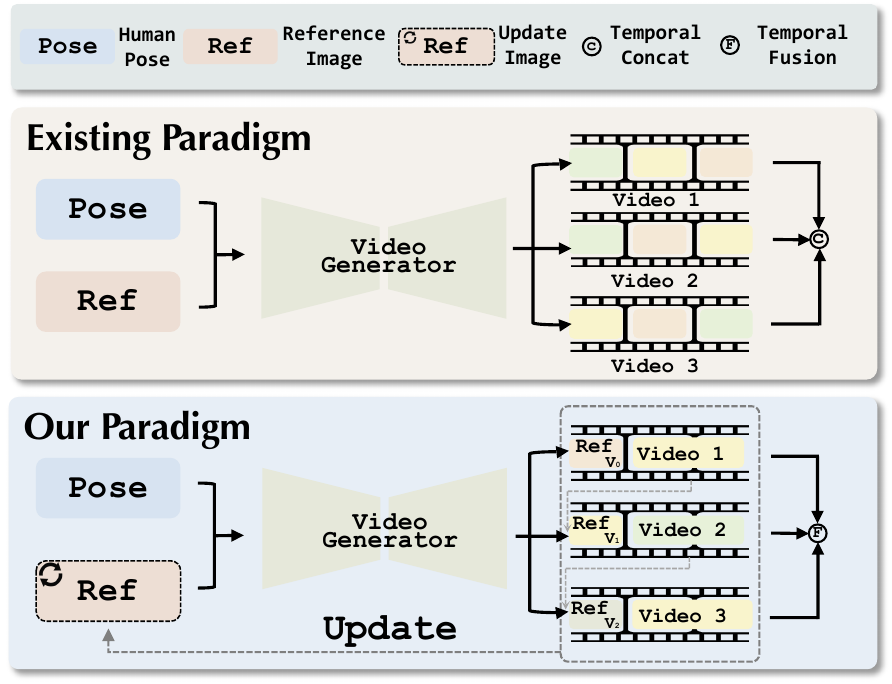}
  \caption{Comparison of generation paradigms. Existing methods (top) rely on a single static reference, leading to independent chunk generation and temporal inconsistencies. Our FrameCache paradigm (bottom) dynamically updates the reference frame via historical generation results, enabling causality-consistent long-sequence animation.}
  \label{fig:fig2}
% \vspace{-1.0em}
\end{figure}
Motivated by these limitations and inspired by human cognition, which flexibly integrates past observations to interpret ongoing actions~\cite{wang2025ignoring}, we introduce FrameCache, a training-free, causality-consistent reference frame framework designed to achieve temporally coherent and visually consistent character animation over long sequences. As conceptually compared in Fig.~\ref{fig:fig2}, instead of relying on a single static image like existing paradigms, FrameCache explicitly converts historical generation results into causal guidance at two complementary levels. First, at the \emph{reference level}, we introduce the Screen-Cache-Match (SCM) strategy to construct a dynamic, high-quality reference memory. It filters frames via a quality-aware mechanism, maintains diversity with a redundancy-aware replacement strategy, and retrieves the most motion-consistent reference for upcoming poses. Second, at the \emph{generative level}, we propose a Trajectory-Aware Autoregressive Generation (TAAG) mechanism that aligns denoising trajectories across adjacent video chunks. By employing an overlap-aware latent propagation and a dual-domain fusion strategy, TAAG seamlessly blends low-frequency structural layouts with high-frequency textural details, preventing abrupt transitions at chunk boundaries.

Compared to existing methods, FrameCache offers several distinct advantages. First, it is entirely training-free and can be seamlessly integrated into diverse diffusion-based baselines as a plug-and-play module. Second, the dual-level consistency—combining semantic appearance anchors from the memory cache with smooth latent trajectory propagation—allows the model to preserve identity and fine-grained details across infinitely long sequences. Third, the framework demonstrates robust generalization, ensuring effective performance across complex motions, severe occlusions, and previously unseen character appearances. In summary, the main contributions of this work are as follows:
\begin{itemize}[leftmargin=1em,topsep=0pt]
\item We propose FrameCache, a causality-consistent, training-free reference frame framework for long-sequence character animation. By operating at both the reference and generative levels, it explicitly converts historical generation results into causal guidance to mitigate temporal flicker, identity drift, and visual artifacts.
\item We design two complementary mechanisms to enable robust long-horizon generation: the Screen-Cache-Match (SCM) strategy that maintains a dynamic reference memory for motion-consistent appearance guidance, and the Trajectory-Aware Autoregressive Generation (TAAG) mechanism that utilizes overlap-aware latent propagation and dual-domain fusion to seamlessly align denoising trajectories.
\item Extensive evaluations demonstrate that FrameCache can be seamlessly integrated into diverse diffusion-based baselines as a plug-and-play module. Notably, it effectively trades negligible low-level feature shifts for substantial, human-perceptible improvements in temporal coherence, identity preservation, and visual stability. Transitioning to the video domain, generative systems face the dual requirement of maintaining high per-frame visual fidelity while ensuring global temporal coherence. 
\end{itemize}

\section{Related Work}
\subsection{Diffusion Models for Image and Video Generation}

Diffusion models have recently emerged as the dominant paradigm in generative modeling, demonstrating exceptional capabilities in fine-grained detail reconstruction and sample diversity control via iterative denoising processes~\cite{ho2020denoising,dhariwal2021diffusion,song2020score,balaji2022ediff,zhang2024motiondiffuse,tevet2022human,wang2025designdiffusion,wu2023harnessing,zhou2024magictailor,shen2023advancing,fei2024diffusion,huang2025attenst,he2024freestyle,yan2025scail,tu2025stableanimator++,tu2025stableavatar,javanmardi2026talkingpose,tu2025motionfollower,cheng2025wan}. Text-guided image generation, in particular, has experienced remarkable advancements. By leveraging powerful language encoders, modern architectures can produce semantically aligned and photorealistic images, establishing a robust foundation for broad cross-modal applications~\cite{saharia2022photorealistic,huang2023composer,ramesh2022hierarchical,mou2024t2i}. Concurrently, open-source frameworks like Stable Diffusion~\cite{zhang2023adding} strike an effective balance between visual quality and computational efficiency, significantly accelerating the deployment of diffusion models in real-world creative design and content generation tasks.

Transitioning to the video domain, generative systems face the dual requirement of maintaining high per-frame visual fidelity while ensuring global temporal coherence. To satisfy such constraints, recent approaches frequently build upon pretrained image diffusion backbones augmented with dedicated temporal modeling modules~\cite{esser2023structure,wu2023tune}. Structural modifications range from injecting pseudo-3D convolutions and spatiotemporal attention mechanisms into U-Net architectures, to adopting scalable Transformer-based designs (DiT) for superior long-range dependency modeling~\cite{wu2023tune,xiang2023versvideo,shi2025decouple}. Furthermore, frameworks like Stable Video Diffusion integrate structural preservation with explicit motion modeling to elevate visual consistency and temporal smoothness~\cite{blattmann2023stable}.

\subsection{Temporal Consistency in Long-Term Character Animation}

Long-term character animation represents a highly demanding subtask within video synthesis, primarily due to the stringent requirements for identity preservation and temporal continuity over extended horizons. While standard diffusion models excel at synthesizing high-quality isolated frames, independent frame-by-frame generation inevitably introduces severe visual artifacts, including flickering, motion discontinuities, and semantic drift. To counteract such degradation, recent literature has explored explicit temporal modeling strategies to align generated frames with driving poses.

Early and concurrent attempts have proposed various architectural adaptations to enforce consistency. For instance, MagicAnimate utilizes a 3D U-Net for joint spatiotemporal modeling~\cite{xu2024magicanimate}, whereas AnimateAnyone integrates a ReferenceNet, a Pose Guider, and temporal attention layers to suppress pose drift and structural inconsistencies~\cite{hu2023animateanyone}. Other works extend these capabilities by introducing 3D parametric guidance (e.g., Champ~\cite{zhu2024champ}), disentangled control mechanisms (e.g., Disco~\cite{wang2024disco}), or confidence-aware pose guidance (e.g., MimicMotion~\cite{zhang2025mimicmotion}). Similarly, UniAnimate evolved from a 3D U-Net backbone into UniAnimate-DiT, adopting a Video Diffusion Transformer with LoRA fine-tuning and a lightweight pose encoder to support extended video generation~\cite{wang2025unianimate}. 

Despite showing considerable promise, prior frameworks are fundamentally bottlenecked by their reliance on a single, static initial frame. When characters undergo complex movements, large rotations, or self-occlusions, a single reference provides insufficient information. This forces the model to hallucinate missing textures, inevitably leading to identity degradation and detail loss over time. Furthermore, due to GPU memory constraints, generating infinitely long videos necessitates autoregressive or chunk-wise generation strategies. Under such paradigms, existing methods lack explicit regularizations to maintain denoising trajectory consistency across segment boundaries. The independent sampling of latent noise for each chunk introduces distribution shifts, culminating in error accumulation, abrupt boundary transitions, and compounding temporal artifacts.

Conversely, our training-free FrameCache framework tackles both shortcomings directly. To overcome the static reference bottleneck, it supplies dynamic, motion-consistent appearance guidance via the Screen-Cache-Match (SCM) strategy, converting ambiguous implicit memory into reliable explicit anchors. Concurrently, to solve chunk boundary discontinuities, it enforces seamless segment transitions through overlap-aware latent trajectory propagation and dual-domain fusion (TAAG). By doing so, FrameCache effectively enhances existing models into causality-consistent long-sequence generators without requiring any structural modifications, computationally expensive training pipelines, or bespoke fine-tuning.

\begin{figure*}[!t]  % <--- 第一处修改：加上感叹号，强制忽略 LaTeX 的“美学留白限制”
  \centering
  % vvv 第二处修改：强行限制图片高度不能超过页面高度的 35%，防止图太高把版面撑爆 vvv
  \includegraphics[width=\linewidth, height=0.8\textheight, keepaspectratio]{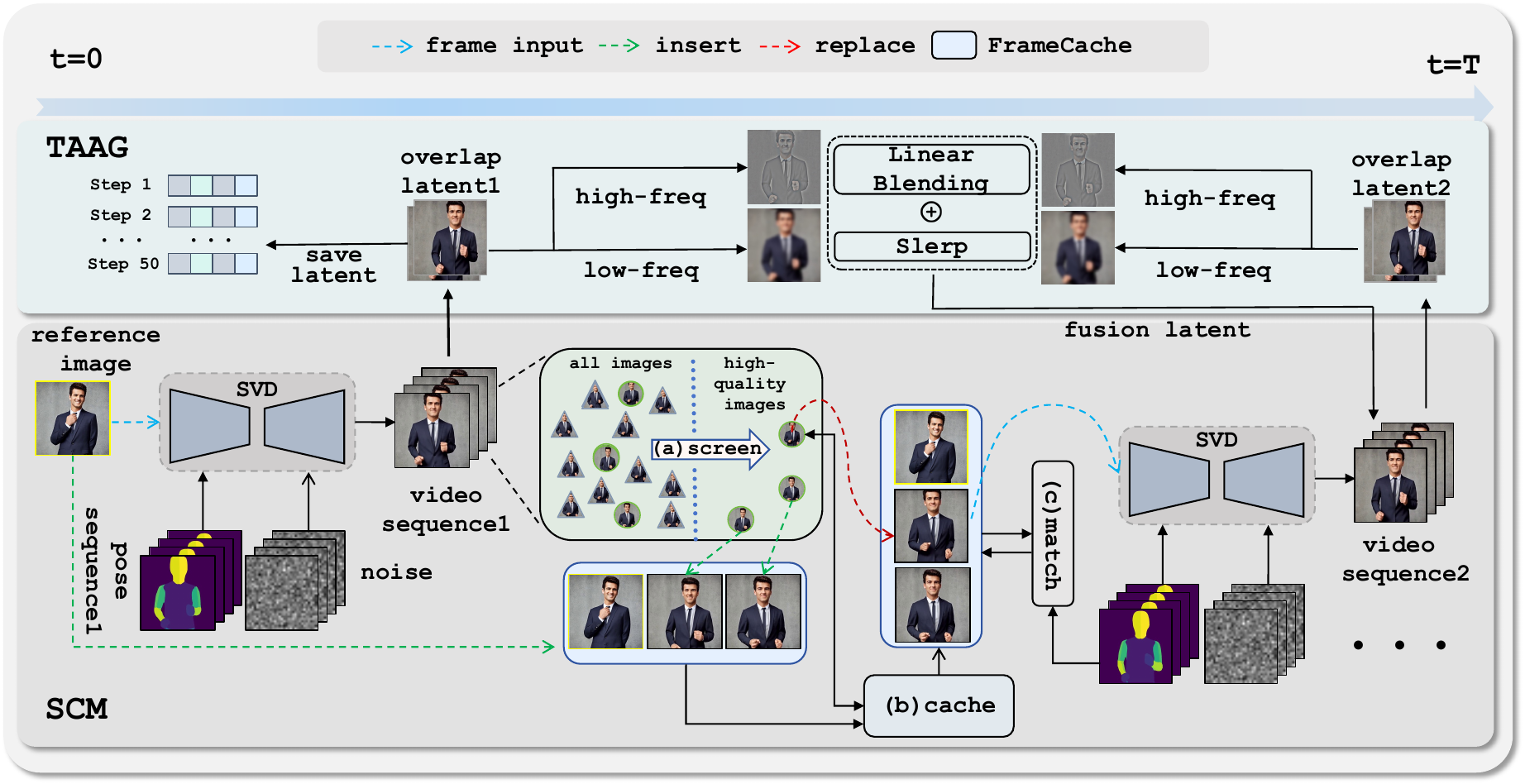}
  \caption{Overview of the proposed \textbf{FrameCache} framework. It achieves causality-consistent generation over long sequences via a dual-level design. \textbf{Bottom:} The Screen-Cache-Match (SCM) mechanism dynamically evaluates generated frames, maintains a diverse reference memory, and retrieves the optimal appearance anchor for the next chunk. \textbf{Top:} The Trajectory-Aware Autoregressive Generation (TAAG) mechanism extracts overlapping latents between adjacent chunks, disentangling them into high- and low-frequency components for specialized fusion, ensuring seamless denoising trajectories.}
  \label{fig:fig3}
\end{figure*}

\section{Method}
We consider the standard pose-guided human animation setting, where a single reference image and a target pose sequence are used to synthesize a video. Let $I_0$ denote the source image and $\mathcal{P}=\{P_t\}_{t=1}^{T}$ denote the driving pose sequence. The objective is to generate an animation $\hat{\mathcal{V}}=\{\hat{I}_t\}_{t=1}^{T}$ that preserves both \emph{temporal coherence} and \emph{appearance consistency} over long horizons.

Despite the strong visual quality of recent diffusion-based animation models, long-sequence generation remains challenging. When the motion becomes large, irregular, or partially occluded, relying on a single fixed reference is often insufficient: the model may gradually lose identity-specific details, exhibit temporal flickering, or produce inconsistent structures across segments. Moreover, in autoregressive or chunk-wise generation, distribution shifts between adjacent segments can accumulate over time, further degrading long-range stability.

To address these issues, we propose \textbf{FrameCache}, a training-free and plug-and-play framework that explicitly converts historical generation results into causal guidance. As illustrated in Fig.~\ref{fig:fig3}, our method achieves causality-consistent synthesis through a dual-level design. At the \emph{reference level} (Sec.~\ref{sec:scm}), the \textbf{Screen-Cache-Match (SCM)} strategy constructs a high-quality, motion-consistent reference memory from previously generated frames. At the \emph{generative level} (Sec.~\ref{sec:taag}), the \textbf{Trajectory-Aware Autoregressive Generation (TAAG)} mechanism ensures that adjacent video chunks remain coherent at both the frame and latent levels via overlap-aware propagation and dual-domain fusion. Together, these components reinterpret long-sequence animation as a \emph{trajectory-conditioned generation problem}, rather than a sequence of isolated frame predictions.

\subsection{Overview}

Given a source image $I_0$ and a long driving pose sequence $\mathcal{P}$, we divide the full sequence into a set of overlapping pose chunks:
\begin{equation}\label{eq:chunks}
\mathcal{P}^{(1)}, \mathcal{P}^{(2)}, \dots, \mathcal{P}^{(K)},
\end{equation}
where each chunk contains $L$ frames and adjacent chunks share $O$ overlapping frames. The overall framework is built on top of an off-the-shelf pose-guided diffusion animation model without requiring any additional training or weight updates.

At a high level, our pipeline proceeds as follows. For the first segment, the frozen diffusion model takes the original source image $I_0$ as appearance guidance and synthesizes the initial chunk. The generated frames are evaluated and selectively inserted into a dynamic cache. For each subsequent segment, instead of conditioning solely on the original source image, the model retrieves a motion-compatible reference from the SCM module. Concurrently, to avoid boundary artifacts, the TAAG module reuses the latent trajectory in the overlapping region. It disentangles the old and new latent states into high- and low-frequency components, applying specialized blending (e.g., Slerp and Linear Blending) to preserve structural and textural continuity. This process continues autoregressively until the full sequence is generated.

\subsection{The Screen-Cache-Match (SCM) Mechanism}
\label{sec:scm}

\begin{figure}[htbp]
  \centering
  % 注意：如果你的图不在 fig 文件夹下，请改成实际路径
  \includegraphics[width=\linewidth]{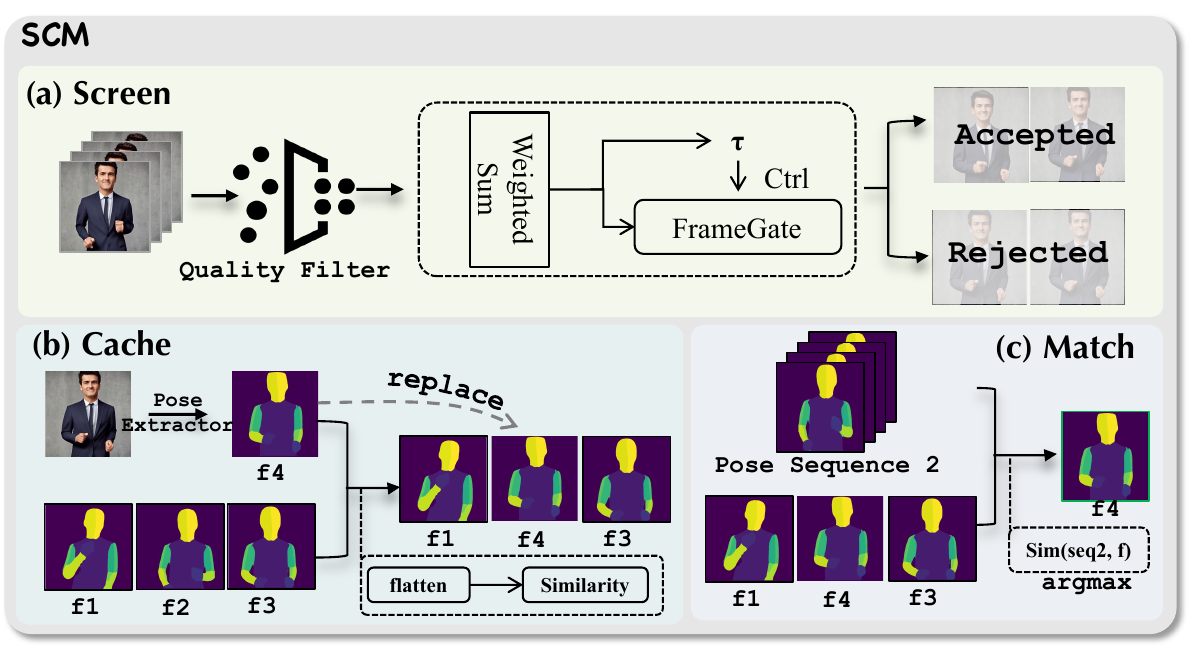}
  \caption{Detailed illustration of the Screen-Cache-Match (SCM) mechanism. (a) \textbf{Screen} filters candidate frames based on perceptual quality scores. (b) \textbf{Cache} maintains a diverse memory buffer by identifying and replacing redundant frames via pose feature similarity. (c) \textbf{Match} retrieves the most motion-compatible reference for the upcoming pose sequence to prevent semantic drift.}
  \label{fig:fig4}
\end{figure}

Operating at the reference level, the SCM module determines \emph{what} visual information should be preserved and \emph{when} it should be reused to minimize identity drift. As detailed in Fig.~\ref{fig:fig4}, the design consists of three sequential stages.

\paragraph{Screen: Quality-aware frame admission.}
In long-sequence generation, not every generated frame is reliable. Admitting frames with blur or structural distortions into the memory would introduce noise and compound errors over time. Inspired by the selectivity refinement mechanism of memory engrams in neuroscience~\cite{2024Dynamic}, we introduce a perceptual quality-driven filtering stage. For each candidate frame $I$, we compute a no-reference perceptual quality score using CLIP-IQA~\cite{wang2023exploring} and MUSIQ~\cite{ke2021musiq}:
\begin{equation}\label{eq:score}
\text{Score}(I)=\lambda Q_{\text{CLIP}}(I)+(1-\lambda)Q_{\text{MUSIQ}}(I),
\end{equation}
where $\lambda$ controls the trade-off between semantic-perceptual and aesthetic quality. To make the admission criterion adaptive, we define a dynamic threshold based on the first accepted frame's score $S_0$:
\begin{equation}\label{eq:tau}
\tau = S_0 \cdot \max\left(\alpha,\frac{1}{1+e^{-2S_0}}\right),
\end{equation}
where $\alpha$ controls the strictness of the filter. A candidate frame is inserted into the cache only if $\text{Score}(I) > \tau$. This ensures the cache remains dominated by visually reliable anchors.

\paragraph{Cache: Redundancy-aware memory maintenance.}
While the Screen stage guarantees quality, it does not guarantee diversity. To prevent redundant samples from monopolizing the memory, we maintain a fixed-capacity reference buffer $\mathcal{C}=\{f_1,f_2,\dots,f_C\}$, updated via a redundancy-aware strategy. Each cached frame $f_i$ is represented by a pose-aware feature tensor $x_i$, yielding a pairwise cosine similarity matrix:
\begin{equation}\label{eq:sim_matrix}
S_{ij}=\frac{x_i^\top x_j}{\|x_i\|\cdot\|x_j\|}.
\end{equation}
The row-wise similarity sum $r_i=\sum_{j\neq i} S_{ij}$ measures how redundant $f_i$ is within the current cache. For a new candidate feature $x_{\text{new}}$ with similarity vector $\mathbf{s}_{\text{new}}$ against cached items, we define the replacement gain $g_i$ as:
\begin{equation}\label{eq:gain}
g_i=\sum_j S_{ij}-2r_i+2\left(\sum_j s_{\text{new},j}-s_{\text{new},i}\right).
\end{equation}
We target the replacement index $i^*=\arg\min_{i>0} g_i$, preserving the first frame ($i=0$) as an immutable global anchor. If the minimal gain satisfies our criterion, $f_{i^*}$ is replaced by $x_{\text{new}}$. Inspired by the pattern-separation effect of PV$^+$ interneurons~\cite{espinoza2018parvalbumin+}, this suppresses redundant activations and maintains a compact, diverse memory covering various viewpoints.

\paragraph{Match: Motion-consistent reference retrieval.}
To prevent semantic drift under large pose transitions, the reference must align with the target motion. For each incoming pose segment, we retrieve the cached frame whose pose representation best matches the target. Let $\mathcal{C}=\{x_1^{\text{ref}},\dots,x_C^{\text{ref}}\}$ denote the cached pose features, and $\{x_t\}_{t=1}^{T'}$ be the pose features of the current target segment. The motion alignment score is:
\begin{equation}\label{eq:match_score}
\text{Sim}(x_i^{\text{ref}})=\frac{1}{T'}\sum_{t=1}^{T'} \frac{{x_i^{\text{ref}}}^\top x_t}{\|x_i^{\text{ref}}\|\cdot\|x_t\|}.
\end{equation}
The optimal retrieved reference is thus:
\begin{equation}\label{eq:match_retrieve}
f^*=\arg\max_{x_i^{\text{ref}}\in\mathcal{C}} \text{Sim}(x_i^{\text{ref}}).
\end{equation}
By replacing a static image with this motion-matched retrieval, SCM provides an adaptive causal prior that substantially reduces temporal flicker.

\subsection{Trajectory-Aware Autoregressive Generation (TAAG)}
\label{sec:taag}

\begin{figure}[htbp]
  \centering
  \includegraphics[width=\linewidth]{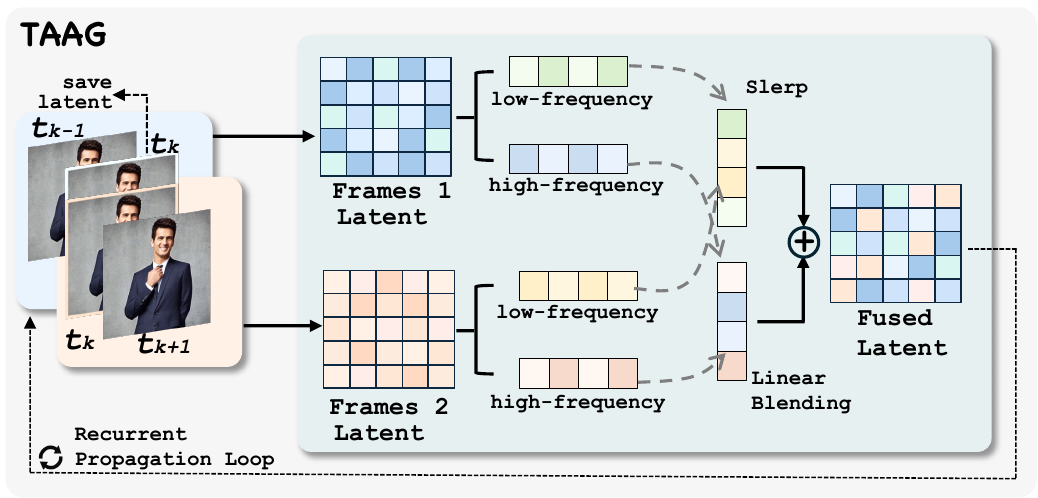}
  \caption{Detailed illustration of the Trajectory-Aware Autoregressive Generation (TAAG) mechanism. Overlapping latents from the previous chunk and the current chunk are disentangled into low-frequency structural components and high-frequency textural details. These components are then adaptively fused to maintain a coherent denoising trajectory across segment boundaries.}
  \label{fig:fig5}
\end{figure}

While SCM determines \emph{which} reference guides a segment, the generative level TAAG mechanism governs \emph{how} adjacent segments connect. As illustrated in Fig.~\ref{fig:fig5}, to avoid discontinuities, we enforce coherence along the denoising trajectory.

\paragraph{Overlap-aware autoregressive segmentation.}
Based on Eq.~\ref{eq:chunks}, the $k$-th chunk is denoted by $\mathcal{P}^{(k)} = \{P_t\}_{t=s_k}^{e_k}$, where the start frame shifts by $s_{k+1}=e_k-O+1$. This overlap $O$ enables shared temporal context. The conditioning image $I_{\text{ref}}^{(k)}$ is dynamically assigned:
\begin{equation}\label{eq:cond_img}
I_{\text{ref}}^{(k)}=
\begin{cases}
I_0, & k=1,\\
\text{Match}(\mathcal{C}, \mathcal{P}^{(k)}), & k>1.
\end{cases}
\end{equation}
This adaptation dynamically alleviates the generative burden on the initial source image.

\paragraph{Latent trajectory propagation.}
Independently denoising overlapping chunks from random noise yields abrupt boundary transitions due to stochasticity. To enforce causal consistency, we initialize and regularize the overlap region's latent states. During the generation of segment $k-1$, we save the latent states corresponding to the overlapping suffix across all denoising steps:
\begin{equation}\label{eq:latent_save}
\mathcal{T}^{(k-1)} = \{ \tilde{z}_{i,\text{overlap}}^{(k-1)} \}_{i=1}^{N},
\end{equation}
where $N$ is the total number of denoising steps. When generating segment $k$, we use $\mathcal{T}^{(k-1)}$ to explicitly regularize the overlapping prefix:
\begin{equation}\label{eq:latent_prop}
\tilde{z}_{i,\text{overlap}}^{(k)} \leftarrow \Phi\!\left(\tilde{z}_{i,\text{overlap}}^{(k-1)},\, \tilde{z}_{i,\text{overlap}}^{(k)}\right),
\end{equation}
where $\Phi$ acts as a trajectory fusion operator. This ensures consistency is baked into the denoising process itself, rather than merely relying on pixel-space blending.

\paragraph{Dual-domain fusion.}
A naive hard overwrite of overlapping latents preserves continuity but suppresses local adaptability. To balance segment-level coherence and local flexibility, we formulate the trajectory fusion operator $\Phi$ as a dual-domain fusion mechanism. Let $\tilde{z}_{i}^{(k-1)}$ and $\tilde{z}_{i}^{(k)}$ denote the previous and current overlap latents at denoising step $i$. We first disentangle these latents into low-frequency and high-frequency components via a 2D Fast Fourier Transform ($\mathcal{F}$). Using a spatial Gaussian low-pass filter mask $M$ with a configurable radius $\sigma$, the low-frequency component $z_{\text{low}}$ and high-frequency component $z_{\text{high}}$ are extracted as:
\begin{equation}
z_{\text{low}} = \mathcal{F}^{-1}(M \odot \mathcal{F}(\tilde{z})), \quad z_{\text{high}} = \mathcal{F}^{-1}((1-M) \odot \mathcal{F}(\tilde{z})),
\end{equation}
where $\odot$ denotes element-wise multiplication and $\mathcal{F}^{-1}$ is the inverse FFT. Intuitively, low-frequency components encode the coarse structural motion layout, while high-frequency components capture sharp, local textural details.

To fuse the low-frequency components while preserving the spherical prior and variance of the diffusion latent space, we employ Spherical Linear Interpolation (Slerp). Given a mixing ratio $\beta \in [0,1]$, the fused low-frequency latent is formulated as:
\begin{equation}\label{eq:slerp}
\hat{z}_{\text{low}}^{(k)} = \frac{\sin((1-\beta)\Omega)}{\sin(\Omega)} z_{\text{low}}^{(k-1)} + \frac{\sin(\beta\Omega)}{\sin(\Omega)} z_{\text{low}}^{(k)},
\end{equation}
where $\Omega$ represents the angle between the two flattened latent vectors, calculated as:
\begin{equation}\label{eq:omega}
\Omega = \arccos\left(\frac{z_{\text{low}}^{(k-1)} \cdot z_{\text{low}}^{(k)}}{\|z_{\text{low}}^{(k-1)}\| \|z_{\text{low}}^{(k)}\|}\right).
\end{equation}
Slerp ensures smooth geometric transitions for the global structure without degrading the latent distribution.

Conversely, for the high-frequency components that represent local details, a standard linear blending is sufficient and ensures textural adaptability. With a high-frequency mixing ratio $\gamma$ (which defaults to $\beta$), the fused high-frequency latent is computed as:
\begin{equation}\label{eq:linear_blend}
\hat{z}_{\text{high}}^{(k)} = (1-\gamma) z_{\text{high}}^{(k-1)} + \gamma z_{\text{high}}^{(k)}.
\end{equation}

Finally, the recombined latent state for the current chunk is obtained by summing the fused components:
\begin{equation}\label{eq:dual_fusion}
\hat{z}_{i}^{(k)} = \hat{z}_{\text{low}}^{(k)} + \hat{z}_{\text{high}}^{(k)}.
\end{equation}
This frequency-aware disentanglement effectively preserves the coarse motion layout inherited from the previous segment while allowing the current segment to refine local details seamlessly.

\paragraph{Unified effect.}
Together, the reference-level SCM and the generative-level TAAG form a cohesive whole. SCM curates and matches the optimal appearance anchor, while TAAG connects adjacent chunks along a consistent denoising trajectory via dual-domain fusion. Consequently, FrameCache seamlessly translates isolated generation into a causality-consistent autoregressive pipeline.

\section{Experimental Evaluations}
In this section, we evaluate the effectiveness of FrameCache by integrating it into three representative human image animation baselines: MagicAnimate~\cite{xu2024magicanimate}, StableAnimator~\cite{tu2025stableanimator}, and DynamiCtrl~\cite{zhao2025dynamictrl}. First, we outline our experimental setup, detailing the dataset, implementation details, and evaluation metrics. Subsequently, we present both quantitative and qualitative results, followed by a comprehensive analysis of our findings.
\subsection{Experimental Setup}
In the experiments, we adhered to the original methods of the respective baselines. For MagicAnimate~\cite{xu2024magicanimate}, we utilized DensePose~\cite{G2018DensePose} to extract the poses, whereas for StableAnimator~\cite{tu2025stableanimator} and DynamiCtrl~\cite{zhao2025dynamictrl}, we employed DWPose~\cite{yang2023effective} to obtain the skeletal poses. To evaluate the methods, we collected 15 high-quality videos from TikTok. During preprocessing, individual video frames were sampled, resized, and center-cropped to match the specific resolution requirements of each baseline: $512 \times 512$ for MagicAnimate and StableAnimator, and $768 \times 1360$ for DynamiCtrl. All experiments were conducted with inference on a single NVIDIA RTX 6000 GPU.

\textbf{Implementation details.} 
For the FrameCache framework, the generation chunk size $L$ and overlap length $O$ are configured according to the distinct temporal contexts of the specific baselines: $L=32$ with $O=4$ for MagicAnimate, $L=16$ with $O=4$ for StableAnimator, and $L=48$ with $O=8$ for DynamiCtrl. In the SCM module, the maximum cache capacity is fixed at $C = 4$. The quality score trade-off parameter is set to $\lambda = 0.5$, and the adaptive admission strictness is $\alpha = 0.95$. For the TAAG mechanism, the spatial Gaussian low-pass filter radius is set to $\sigma = 8.0$. The mixing ratios for both low-frequency Slerp and high-frequency linear blending are empirically set to $\beta = 0.5$ and $\gamma = 0.5$, respectively. Except for the baseline-specific sequence chunking ($L$ and $O$), all other algorithmic hyperparameters are kept strictly constant across all evaluated baselines, effectively demonstrating the plug-and-play robustness of our proposed method.

\subsection{Quantitative and Qualitative Results}
\textbf{Evaluation metrics.}
Our evaluation adopts a comprehensive suite of no-reference (NR) metrics to assess perceptual realism and aesthetic quality without relying on ground-truth references. Specifically, we include CNN-IQA~\cite{kang2014convolutional}, DBCNN~\cite{zhang2020blind}, MUSIQ~\cite{ke2021musiq}, NIMA~\cite{talebi2018nima}, PAQ2PIQ~\cite{ying2020patches}, TRES-FLIVE~\cite{golestaneh2022no}, and WaDIQaM-NR~\cite{bosse2018deep}. For all of these selected metrics, a higher score indicates superior visual quality.

\begin{table*}[htbp]
    \centering
    \caption{Results of Quantitative Comparison. $\uparrow$ indicates that a larger value is preferred. Best results for each baseline are highlighted in \textbf{bold}.}
    \label{tab:framecache_results}
    \renewcommand{\arraystretch}{1.2}
    \setlength{\tabcolsep}{4.5pt}
    \begin{tabular}{llccccccc}
    \toprule
    \textbf{Method} & \textbf{Setting} & \textbf{CNN-IQA}$\uparrow$ & \textbf{DBCNN}$\uparrow$ & \textbf{MUSIQ}$\uparrow$ & \textbf{NIMA}$\uparrow$ & \textbf{PAQ2PIQ}$\uparrow$ & \textbf{TRES-FLIVE}$\uparrow$ & \textbf{WaDIQaM-NR}$\uparrow$ \\ 
    \midrule
    \multirow{2}{*}{\textbf{MagicAnimate}~\cite{xu2024magicanimate}} 
    & Origin     & 0.434 & 0.352 & \textbf{54.1} & 4.63 & 68.2 & \textbf{83.7} & -0.587 \\
    & FrameCache & \textbf{0.438} & \textbf{0.353} & \textbf{54.1} & \textbf{4.87} & \textbf{68.5} & 83.6 & \textbf{-0.586} \\ 
    \midrule
    \multirow{2}{*}{\textbf{StableAnimator}~\cite{tu2025stableanimator}} 
    & Origin     & 0.544 & 0.467 & 64.7 & 5.49 & 73.3 & 87.6 & -0.374 \\
    & FrameCache & \textbf{0.549} & \textbf{0.473} & \textbf{65.1} & \textbf{5.55} & \textbf{73.8} & \textbf{88.2} & \textbf{-0.368} \\ 
    \midrule
    \multirow{2}{*}{\textbf{DynamiCtrl}~\cite{zhao2025dynamictrl}} 
    & Origin     & 0.600 & 0.490 & 64.2 & 5.06 & 71.1 & 86.5 & -0.262 \\
    & FrameCache & \textbf{0.611} & \textbf{0.502} & \textbf{64.5} & \textbf{5.20} & \textbf{71.3} & \textbf{86.8} & \textbf{-0.235} \\ 
    \bottomrule
    \end{tabular}
\end{table*}

\textbf{Quantitative results.}
As shown in Table~\ref{tab:framecache_results}, FrameCache achieves comprehensive improvements across diverse baselines, confirming its efficacy in elevating per-frame image quality. Notably, when integrated into StableAnimator and DynamiCtrl, our framework consistently outperforms the original methods across all evaluated no-reference image quality metrics. For instance, the improvements in metrics like MUSIQ and NIMA indicate enhanced aesthetic appeal and perceptual realism, while gains in CNN-IQA and DBCNN suggest a noticeable reduction in structural artifacts such as blurriness or noise. For MagicAnimate, the performance remains highly competitive; FrameCache secures gains in the majority of metrics (e.g., PAQ2PIQ and WaDIQaM-NR) and maintains stable behavior elsewhere without introducing degradation. Overall, these results confirm the broad applicability of FrameCache as a training-free, plug-and-play module. It should be emphasized, however, that the core contribution of our causality-consistent framework—suppressing temporal flicker, semantic jitter, and identity drift over long sequences—is inherently a spatio-temporal challenge. Static frame-level metrics, while useful for measuring spatial fidelity, do not fully capture cross-frame coherence. Consequently, the true magnitude of our improvements is much more evident in the qualitative visual comparisons presented next.

% \begin{figure*}[ht]
%   \centering
%   \includegraphics[width=0.9\linewidth]{fig/Qualitative_image_0917.jpg}
%   \caption{Results of qualitative comparison, highlighting the regions enclosed in red boxes and the inter-frame inconsistencies.
% }
%   \label{fig:Qualitative_image}
% \end{figure*}

\begin{figure}[htbp]
  \centering
  \includegraphics[width=\linewidth]{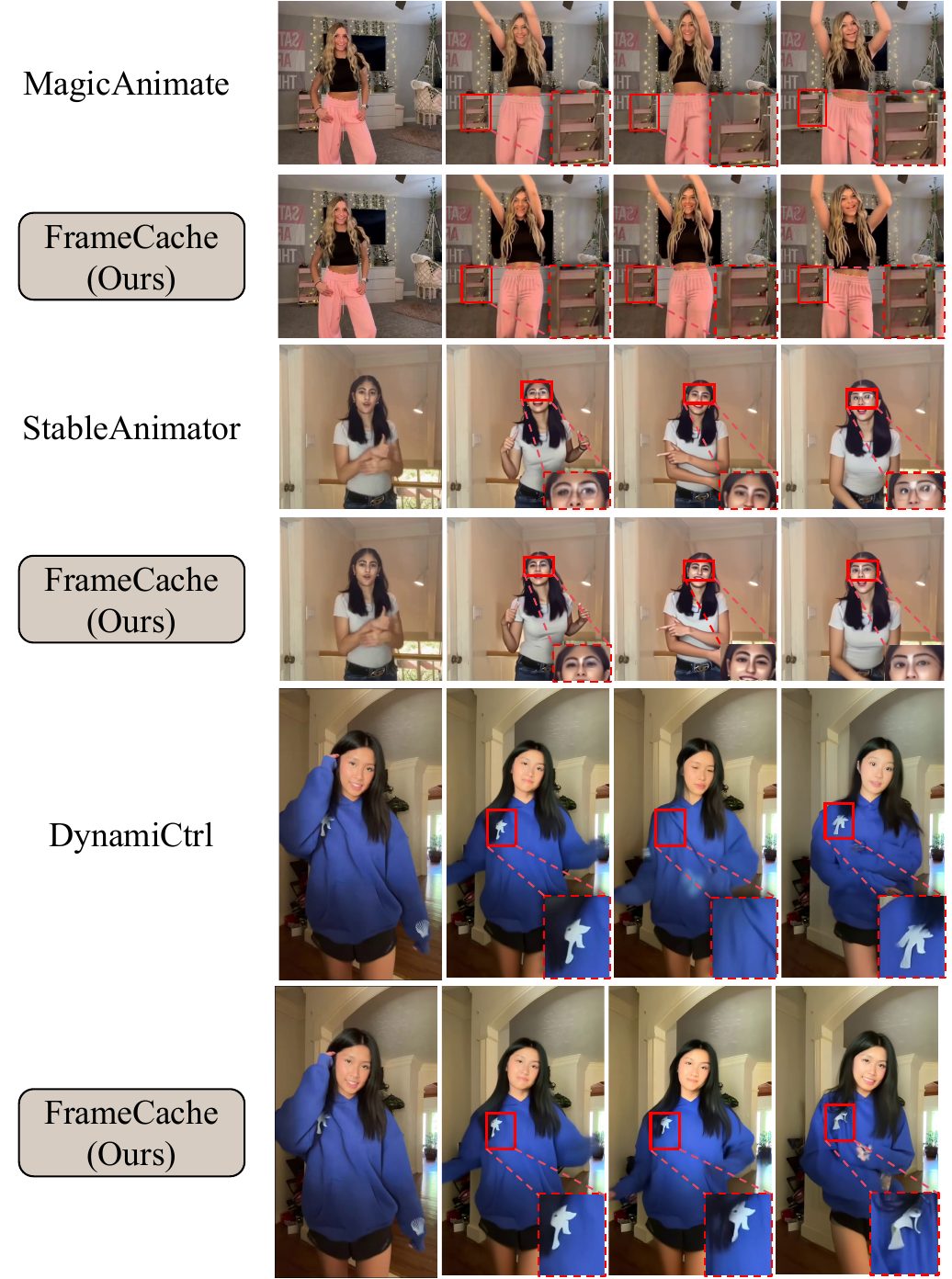}
  \caption{Qualitative comparison of long-sequence generation across three baselines (MagicAnimate, StableAnimator, and DynamiCtrl). The red bounding boxes highlight regions where our FrameCache framework significantly eliminates temporal inconsistencies and artifacts compared to the original methods, with specific improvements observed in background stability, facial identity preservation, and clothing structures.}
  \label{fig:fig6}
\end{figure}

\textbf{Qualitative results.}
As illustrated in Fig.~\ref{fig:fig6}, we conduct a comprehensive qualitative comparison between the original baselines and our FrameCache-enhanced pipelines. It is worth noting that all test videos are unseen during model training and contain complex, large-scale motions that typically challenge generative priors. We analyze the improvements across three critical dimensions: background stability, facial appearance, and clothing consistency. 

First, regarding \emph{background stability}, original diffusion-based methods often suffer from background shifting or local warping as the foreground character moves, a common symptom of independently denoising isolated chunks. FrameCache effectively suppresses these hallucinated background variations by maintaining a coherent denoising trajectory via our TAAG mechanism. Second, in terms of \emph{facial appearance}, preserving character identity over long sequences is notoriously difficult. While baselines tend to exhibit facial distortions, smoothing, or severe identity drift under extreme head rotations and temporal error accumulation, our framework ensures that facial characteristics remain stable, sharp, and highly recognizable across all frames. Third, for \emph{clothing and accessories}, the red bounding boxes in Fig.~\ref{fig:fig6} reveal that original methods struggle to maintain consistent topologies for garments (e.g., jacket hems, exposed waistbands, or printed logos) when previously occluded regions become visible. Because they rely heavily on a single static reference, such implicit structural information is easily lost or inconsistently hallucinated as the pose deviates significantly from the initial frame. In contrast, FrameCache dynamically retrieves motion-aligned references via the SCM module, transforming ambiguous implicit memory into explicit visual cues. By seamlessly fusing these dynamic anchors with overlap-aware latent propagation, our method effectively prevents the sudden appearance or disappearance of fine-grained details. Overall, these visual comparisons highlight the undeniable superiority of FrameCache in preserving high-fidelity identity and producing vivid, temporally seamless animations under challenging conditions.

\begin{table*}[!t]
\centering
\caption{Results of the User Study. We compare FrameCache-enhanced models against their original baselines. The percentage columns report the voting distribution as FrameCache Better (B) / Equal (C) / Baseline Better (A). The Suppression Ratio (SR) is defined as $(B + C) / (A + C)$, where a value $>1$ indicates the superiority of our method.}
\label{tab:user_study}
\resizebox{\textwidth}{!}{
\begin{tabular}{l | c c c | c c c}
\toprule
\multirow{2}{*}{\textbf{Method}} & \multicolumn{3}{c|}{\textbf{Voting Percentage (B / C / A) (\%) $\uparrow$}} & \multicolumn{3}{c}{\textbf{Suppression Ratio (SR) $\uparrow$}} \\
\cmidrule(lr){2-4} \cmidrule(lr){5-7}
% ======= 核心修改在这里：用 \makebox[2cm][c]{...} 强行把右边三个表头撑开到 2cm 宽 =======
& \textbf{TC} & \textbf{IAP} & \textbf{OVQ} & \makebox[2cm][c]{\textbf{TC}} & \makebox[2cm][c]{\textbf{IAP}} & \makebox[2cm][c]{\textbf{OVQ}} \\
\midrule
MagicAnimate~[38]   & 39.5 / 38.9 / 21.6 & 41.9 / 41.1 / 17.1 & 42.1 / 35.2 / 22.7 & 1.30 & 1.43 & 1.34 \\
StableAnimator~[29] & 39.5 / 40.8 / 19.7 & 45.6 / 33.1 / 21.3 & 40.5 / 39.2 / 20.3 & 1.33 & 1.45 & 1.34 \\
DynamiCtrl~[44]     & 41.9 / 36.3 / 21.9 & 42.7 / 37.6 / 19.7 & 41.9 / 36.0 / 22.1 & 1.34 & 1.40 & 1.34 \\
\bottomrule
\end{tabular}
}
\end{table*}

\subsection{User Study}
 
To comprehensively evaluate the perceptual quality and temporal coherence of the generated videos in a manner that aligns with human perception, we conducted a subjective user study involving 25 participants. The study employed a blind A/B testing approach. Participants were presented with side-by-side video pairs generated by the original baseline methods and our FrameCache-enhanced counterparts. The left-right order of the videos was strictly randomized to avoid positional bias. 

Participants were instructed to evaluate the generation results based on three core dimensions: 
(1) \textbf{Temporal Consistency (TC):} the absence of flickering, sudden jumps, or unsmooth motion transitions; 
(2) \textbf{Identity and Appearance Preservation (IAP):} the consistency of fine-grained details (e.g., waistbands, accessories, and clothing topology) throughout the long sequence without distortion or loss; 
(3) \textbf{Overall Visual Quality (OVQ):} the clarity, realism, and severity of artifacts in the generated frames.

For each comparison pair, participants voted for one of three options: "Baseline is better" ($A$), "FrameCache is better" ($B$), or "Equal" ($C$). To quantitatively assess the robust superiority of our framework, we introduce the \textbf{Suppression Ratio (SR)}, defined as $(B + C) / (A + C)$. This metric effectively measures the likelihood of FrameCache performing equally or better compared to the baseline doing the same, explicitly penalizing noticeable degradations.

As reported in Table~\ref{tab:user_study}, FrameCache demonstrates consistent and robust improvements across all three baselines. Across all evaluations, the preference rate for FrameCache ($B$, typically ranging from 39.5\% to 45.6\%) is substantially higher than that for the original baselines ($A$, strictly below 23\%). All Suppression Ratios exceed 1.30, verifying the generalized advantage of our method. Notably, our framework excels in Identity and Appearance Preservation (IAP), achieving the highest SR of 1.45 when integrated into StableAnimator and 1.43 with MagicAnimate. This indicates a significant enhancement in maintaining complex reference details under large motions. Furthermore, the combination of high "Equal" ($C$) and "FrameCache better" ($B$) votes confirms that our training-free SCM and TAAG mechanisms successfully suppress temporal drift and identity degradation while perfectly retaining the base generative capabilities of the original models.

\textbf{Ablation studies.} 
Due to space limitations, comprehensive ablation studies—including evaluations of the individual contributions of the SCM and TAAG modules, as well as analyses of hyperparameter sensitivity—are detailed in the supplementary material.

\section{Conclusion}

In this work, we present FrameCache, a training-free, causality-consistent framework tailored for long-term character animation. By integrating the Screen-Cache-Match (SCM) strategy with Trajectory-Aware Autoregressive Generation (TAAG), FrameCache effectively converts implicit temporal cues into explicit visual and latent guidance. This dual-level design successfully mitigates temporal flickering, identity drift, and structural artifacts inherent in chunk-wise generation. As a lightweight, plug-and-play module, FrameCache significantly enhances existing diffusion baselines, paving the way for robust, reliable, and infinite-length animation in real-world applications such as robotics and human-robot interaction.

% ==================== 参考文献 ====================
\bibliographystyle{ACM-Reference-Format}
\bibliography{acmart} % 对应你的 ref.bib

\appendix

\clearpage % <=== 关键命令：强制把后面的内容挤到全新的一页

% 把附录标题和你的大表格一起放进 strip 环境里，强行横跨双栏！
\begin{strip}
    \vspace*{-1.0cm} % <=== 关键 1：强制向上拉，吃掉 strip 默认的顶部大白边
    % --- 手动伪造的附录大标题 ---
    \begin{center}
        \huge \textbf{Appendix} \\[10pt]
    \end{center}
    \vspace{0.8cm} % 标题和表格之间留出呼吸空间
% \end{strip}

% \begin{strip}
    \vspace{-0.75cm} % 稍微向上收一点边距，让排版更紧凑
    \centering
    \captionof{table}{Quantitative ablation study of the proposed components. The evaluation is conducted on the StableAnimator~\cite{tu2025stableanimator} baseline. Best results across all configurations are highlighted in \textbf{bold}.}
    \label{tab:ablation}
    \renewcommand{\arraystretch}{1.2}
    \setlength{\tabcolsep}{6pt}
    \begin{tabular}{lccccccc}
    \toprule
    \textbf{Model Configuration} & \textbf{CNN-IQA}$\uparrow$ & \textbf{DBCNN}$\uparrow$ & \textbf{MUSIQ}$\uparrow$ & \textbf{NIMA}$\uparrow$ & \textbf{PAQ2PIQ}$\uparrow$ & \textbf{TRES-FLIVE}$\uparrow$ & \textbf{WaDIQaM-NR}$\uparrow$ \\ 
    \midrule
    Baseline              & 0.544 & 0.467 & 64.7 & 5.49 & 73.3 & 87.6 & -0.374 \\
    Baseline + SCM        & 0.542 & 0.469 & 65.2 & 5.29 & \textbf{73.8} & 88.1 & -0.378 \\
    Baseline + TAAG       & 0.547 & \textbf{0.474} & \textbf{65.4} & 5.24 & 73.5 & 87.8 & \textbf{-0.368} \\
    FrameCache (Ours)     & \textbf{0.549} & 0.473 & 65.1 & \textbf{5.55} & \textbf{73.8} & \textbf{88.2} & \textbf{-0.368} \\
    \bottomrule
    \end{tabular}
     % 底部留出与正文的呼吸空间
\end{strip}

\section{Quantitative Ablation Study}
\begin{figure*}[t]
  \centering
  % 注意：请把你的新图命名为 fig8.pdf 或 fig8.jpg 并放到 fig 文件夹下
  \includegraphics[width=\linewidth]{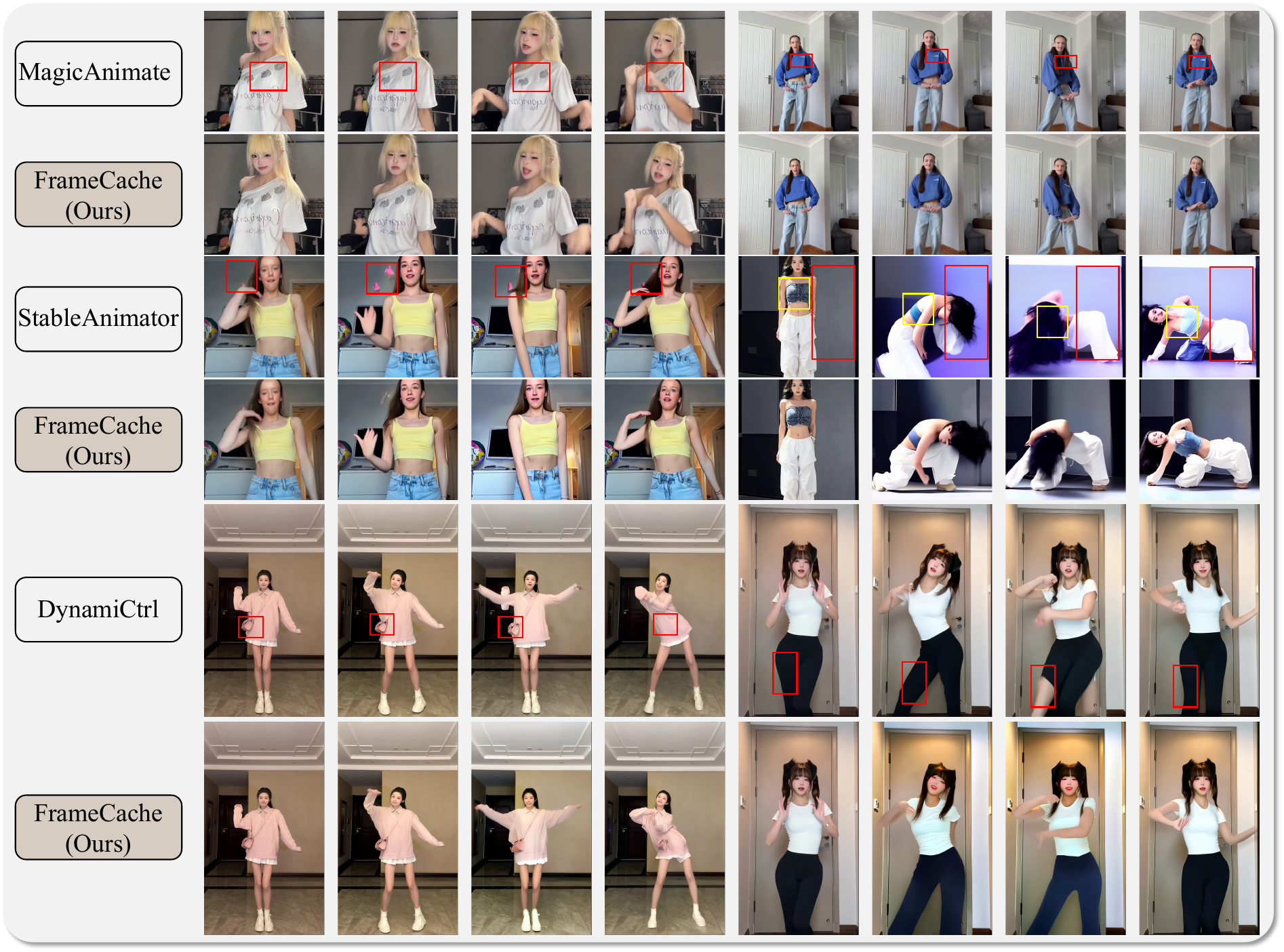}
  \caption{Extended qualitative comparison of long-term identity preservation. While baselines suffer from progressive semantic drift and detail degradation (red boxes), FrameCache (via TAAG) consistently maintains high-fidelity structural and textural details throughout the entire sequence.}
  \label{fig:fig8}
\end{figure*}
To validate the individual contributions of our proposed components, we conduct a comprehensive quantitative ablation study. We select StableAnimator~\cite{tu2025stableanimator} as the representative baseline due to its rigorous temporal constraints. The evaluation incrementally integrates the Screen-Cache-Match (SCM) module and the Trajectory-Aware Autoregressive Generation (TAAG) mechanism. The quantitative results are summarized in Table~\ref{tab:ablation}.

\textbf{Effectiveness of the SCM Module.} 
When the SCM module is integrated independently (\textit{Baseline + SCM}), we observe notable improvements in structural and quality-aware metrics such as PAQ2PIQ and TRES-FLIVE. This confirms that explicitly updating the appearance anchor via motion-consistent retrieval successfully mitigates identity hallucination and provides richer semantic details than a single static reference. However, without latent trajectory alignment, a naive substitution of the conditioning image inevitably introduces slight visual discontinuities at the boundaries of adjacent video chunks. Consequently, aesthetic and noise-sensitive metrics like NIMA and CNN-IQA experience a slight drop compared to the baseline, which quantitatively reflects this boundary jitter.

\textbf{Effectiveness of the TAAG Module.} 
Conversely, applying only the TAAG mechanism (\textit{Baseline + TAAG}) explicitly regularizes the denoising trajectory. By fusing low-frequency and high-frequency latents in overlapping regions, it effectively smooths out temporal noise and eliminates inter-chunk flickering, leading to gains in metrics like CNN-IQA, DBCNN, and MUSIQ. Nevertheless, because it still relies entirely on the initial static frame for appearance guidance, the model suffers from semantic ambiguity and detail loss as the motion deviates significantly from the initial pose, resulting in a suboptimal NIMA score.

\textbf{Synergy of the Full Framework.} 
The full \textit{FrameCache} pipeline yields the most balanced and optimal performance. The SCM module provides accurate, dynamic semantic anchors, while the TAAG mechanism ensures these updated anchors are integrated seamlessly without causing latent trajectory fractures. This synergy perfectly compensates for the individual shortcomings of each module—recovering the drop in NIMA to achieve the highest score of 5.55, while simultaneously securing peak performance in CNN-IQA and TRES-FLIVE. This demonstrates that operating jointly at both the reference and generative levels is indispensable for causality-consistent human animation.
% 【修改点：把表格代码提前到章节标题上方，并使用 [t] 强制置顶】
\begin{table}[t]
    \centering
    \caption{Computational overhead comparison. We report the generation speed in Frames Per Second (FPS) and the Peak Memory Allocated in Gigabytes (GB). The evaluation is conducted on a single NVIDIA RTX 6000 GPU.}
    \label{tab:overhead}
    \renewcommand{\arraystretch}{1.2}
    \setlength{\tabcolsep}{5pt}
    \begin{tabular}{lcccc}
    \toprule
    \multirow{2}{*}{\textbf{Method}} & \multicolumn{2}{c}{\textbf{Speed (FPS) $\uparrow$}} & \multicolumn{2}{c}{\textbf{Peak VRAM (GB) $\downarrow$}} \\
    \cmidrule(lr){2-3} \cmidrule(lr){4-5}
    & \textbf{Baseline} & \textbf{Ours} & \textbf{Baseline} & \textbf{Ours} \\
    \midrule
    MagicAnimate~\cite{xu2024magicanimate}   & 0.794 & 0.398 & 17.6 & 15.5 \\
    StableAnimator~\cite{tu2025stableanimator} & 0.670 & 0.457 & 6.58 & 6.45 \\
    DynamiCtrl~\cite{zhao2025dynamictrl}     & 0.180 & 0.108 & 24.7 & 29.9 \\
    \bottomrule
    \end{tabular}
\end{table}

\vspace{-0.1cm}
\section{Hyperparameter Sensitivity: Overlap Length}

\begin{table*}[t]
    \centering
    \caption{Hyperparameter sensitivity analysis of the overlap length ($O$) within the TAAG mechanism. The evaluation is conducted on the StableAnimator baseline. Best results are highlighted in \textbf{bold}.}
    \label{tab:overlap_ablation}
    \renewcommand{\arraystretch}{1.2}
    \setlength{\tabcolsep}{8pt} % 这里稍微调大了列间距，让通栏表格更舒展、美观
    \begin{tabular}{cccccccc}
    \toprule
    \textbf{Overlap ($O$)} & \textbf{CNN-IQA}$\uparrow$ & \textbf{DBCNN}$\uparrow$ & \textbf{MUSIQ}$\uparrow$ & \textbf{NIMA}$\uparrow$ & \textbf{PAQ2PIQ}$\uparrow$ & \textbf{TRES-FLIVE}$\uparrow$ & \textbf{WaDIQaM-NR}$\uparrow$ \\ 
    \midrule
    $1$ & 0.537 & 0.466 & 64.6 & 5.27 & \textbf{73.8} & 88.0 & -0.392 \\
    $2$ & 0.537 & 0.465 & 64.6 & 5.27 & \textbf{73.8} & 88.1 & -0.390 \\
    $4$ & \textbf{0.549} & \textbf{0.473} & \textbf{65.1} & \textbf{5.55} & \textbf{73.8} & \textbf{88.2} & \textbf{-0.368} \\
    $8$ & 0.542 & 0.467 & 64.6 & 5.28 & \textbf{73.8} & 88.1 & -0.387 \\
    \bottomrule
    \end{tabular}
\end{table*}

To further investigate the impact of the temporal context window, we analyze the sensitivity of the overlap length parameter ($O$) within the TAAG mechanism. As presented in Table~\ref{tab:overlap_ablation}, we evaluate the performance of the StableAnimator baseline under different overlap lengths ($O \in \{1, 2, 4, 8\}$).

The results reveal a clear performance peak at $O=4$. When the overlap length is too small ($O \le 2$), the temporal context is insufficient to bridge the latent distribution shift between adjacent chunks, resulting in suboptimal denoising alignment and lower perceptual quality scores (e.g., NIMA drops to 5.27). Conversely, an excessively large overlap ($O=8$) forces the model to heavily reuse historical latents, which over-constrains the generative flexibility for new motions and dilutes sharp textural details, causing a slight degradation in metrics like CNN-IQA and MUSIQ compared to the peak values. Therefore, $O=4$ provides the optimal trade-off, ensuring seamless boundary transitions while preserving fine-grained visual fidelity.

\section{Computational Overhead Analysis}

\begin{figure*}[t]
  \centering
  % 请确保将你的新图命名为 fig9.pdf 或 fig9.jpg 并放入 fig 文件夹
  \includegraphics[width=\linewidth]{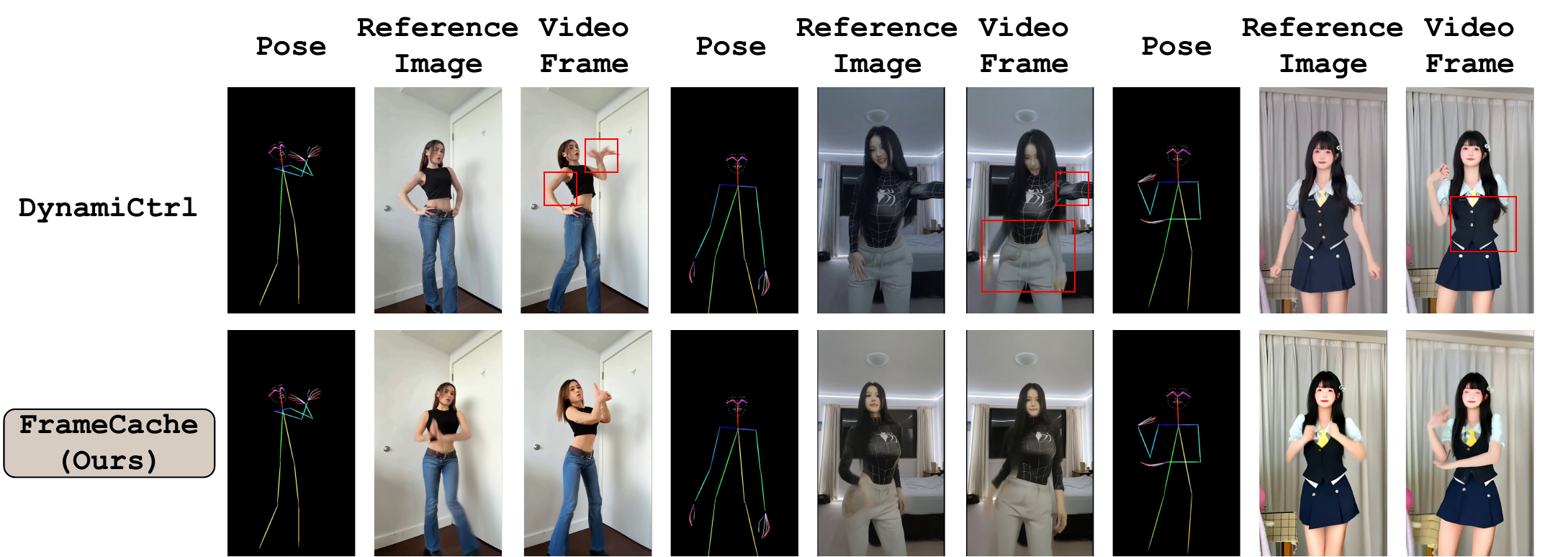}
\caption{Visual comparison of pose fidelity and physical plausibility. Large pose discrepancies force the baseline to hallucinate severe anatomical artifacts like extra limbs (top). By retrieving motion-matched references, FrameCache synthesizes physically accurate frames (bottom).}
  \label{fig:fig9}
  \vspace{-0.3cm}
\end{figure*}

As a plug-and-play framework, FrameCache introduces additional operations during the inference phase, primarily from the quality-aware screening in the SCM module and the dual-domain latent fusion in the TAAG mechanism. To provide a transparent assessment, we evaluate both the temporal (Frames Per Second, FPS) and spatial (Peak VRAM Allocated) overhead of our method compared to the original baselines, averaged over $21$ generated video samples. The quantitative results are summarized in Table~\ref{tab:overhead}.

\textbf{Inference Time Overhead.} 
On a single NVIDIA RTX 6000 GPU, integrating FrameCache yields an inference speed of approximately $50\%$ to $68\%$ of the baseline's FPS. This translates to an overall inference time multiplier of roughly $1.5\times$ to $2.0\times$ compared to the standard chunk-wise generation. This moderate increase is mainly attributed to the explicit feature matching and filtering steps in the SCM module, as well as the Fast Fourier Transform (FFT) and Spherical Linear Interpolation (Slerp) operations required by TAAG.

\textbf{VRAM Footprint.} 
Regarding spatial complexity, we report the Peak Memory Allocated to reflect the true algorithmic footprint. As shown in Table~\ref{tab:overhead}, FrameCache introduces only a marginal VRAM overhead (e.g., an increase of $5.2$ GB for DynamiCtrl) and remarkably even slightly reduces the peak allocation for MagicAnimate and StableAnimator. This optimized memory footprint is due to the efficient memory management of the dynamic cache ($C=4$) and the localized processing of overlapping latents. Crucially, owing to our autoregressive chunk-wise design, this peak VRAM requirement remains strictly \emph{constant} regardless of the total generated video length, ensuring that infinite-length animation remains feasible without memory explosion.

\textbf{Favorable Trade-off.} 
Importantly, we argue that the $1.5\times \sim 2.0\times$ time overhead and negligible VRAM variation represent a highly favorable trade-off. Achieving comparable long-term temporal consistency conventionally requires end-to-end spatiotemporal fine-tuning or training bespoke Video Diffusion Models, which demands hundreds of GPU hours and massive memory clusters. In stark contrast, FrameCache is entirely \emph{training-free}. These modest inference costs elegantly bypass the prohibitively expensive training phase, while simultaneously unlocking robust, infinite-length generation capabilities with strict physical and identity preservation.

\section{Additional Qualitative Results}
\begin{figure*}[htbp]
  \centering
  % 请确保将你的新图命名为 fig10.pdf 或 fig10.jpg 并放入 fig 文件夹
  \includegraphics[width=\linewidth]{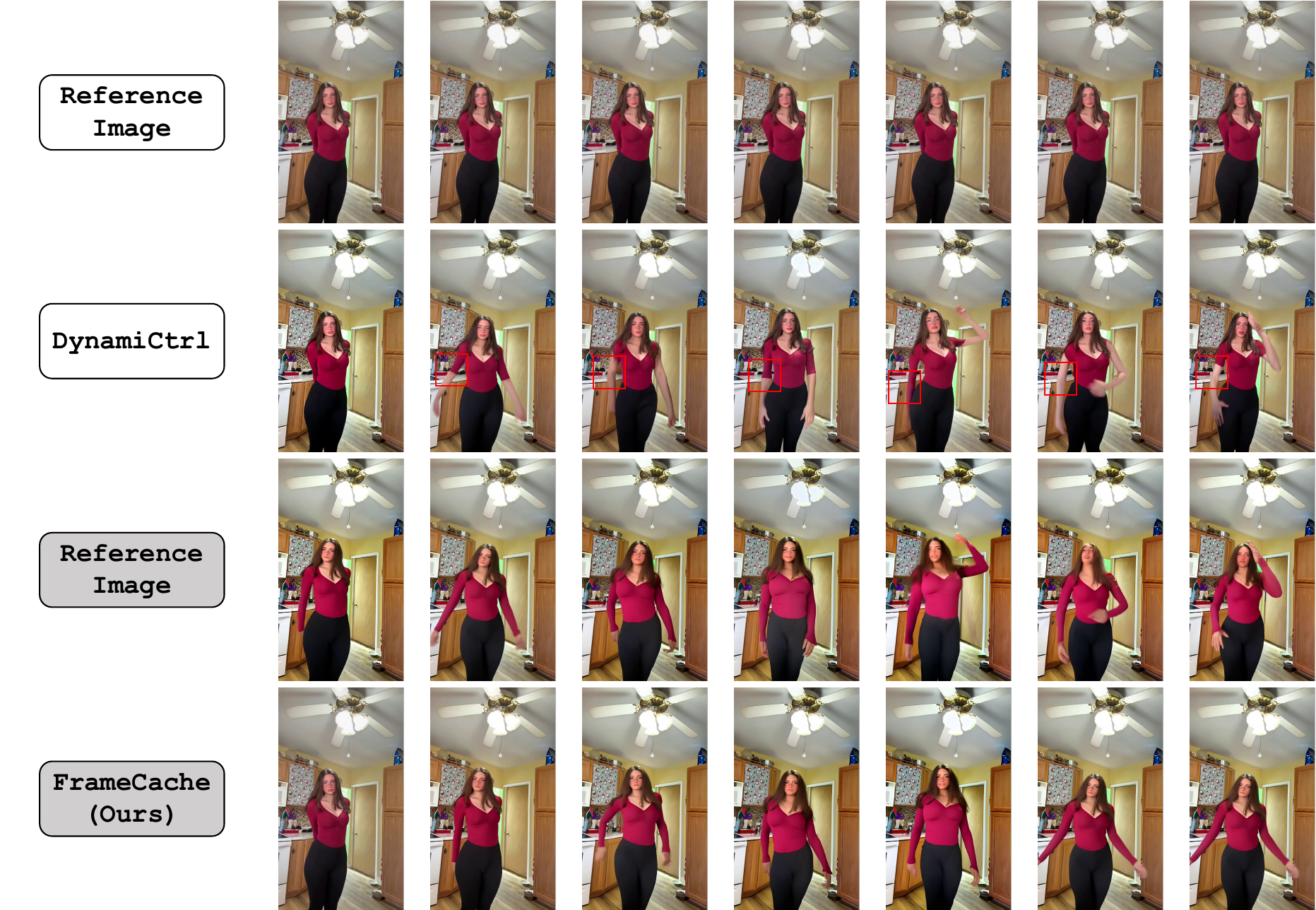}
\caption{Case study of the SCM working mechanism. The baseline struggles with ambiguous references, causing sleeves to inconsistently flicker (top, red boxes). FrameCache retrieves topologically explicit cached frames (bottom) to preserve the correct garment structure seamlessly.}
  \label{fig:fig10}
\end{figure*}
\subsection{Impact of Motion-Consistent Reference Retrieval}

To further demonstrate the robust generative capabilities of our framework over extended temporal horizons, we provide additional visual comparisons in Fig.~\ref{fig:fig8}. 

While the main manuscript highlights the ability of the SCM module to explicitly maintain topological consistency when previously unseen or occluded regions become visible, this section emphasizes the critical role of the \textbf{Trajectory-Aware Autoregressive Generation (TAAG)} mechanism in preventing long-term semantic drift. 

As the sequence generation progresses far beyond the initial frame, standard autoregressive or chunk-wise baselines (e.g., MagicAnimate~\cite{xu2024magicanimate}, StableAnimator~\cite{tu2025stableanimator}, and DynamiCtrl~\cite{zhao2025dynamictrl}) inevitably accumulate latent noise and distribution shifts. Visually, this manifests as a gradual degradation of the character's identity, blurring of clothing logos, and distortion of garment structures in later frames. In contrast, FrameCache explicitly regularizes the latent denoising trajectory across overlapping chunks via TAAG. As shown in the red bounding boxes in Fig.~\ref{fig:fig8}, even under complex and prolonged motions, our method flawlessly preserves the fine-grained details and structural integrity of the original source image from the beginning to the end of the sequence. This confirms that FrameCache not only resolves short-term boundary jitter but also fundamentally stabilizes long-range generation.

To further emphasize the necessity of the Screen-Cache-Match (SCM) module, we present an additional qualitative analysis in Fig.~\ref{fig:fig9}, using DynamiCtrl as the representative baseline. 

A fundamental limitation of existing paradigms is their reliance on a single, static reference image. When the target driving pose exhibits a massive spatial or angular deviation from this fixed reference, the generative model faces an excessively high modeling difficulty. It struggles to manually bridge this massive semantic gap, frequently leading to poor pose-to-appearance alignment. As highlighted by the red bounding boxes in the top row of Fig.~\ref{fig:fig9}, this forced hallucination causes the baseline to produce physically counterintuitive and grotesque artifacts, such as generating two left hands (first example) or hallucinating a third arm (second example).

In stark contrast, our FrameCache framework explicitly mitigates this issue. By continuously updating the memory cache and retrieving the most motion-compatible reference for the upcoming segment, FrameCache ensures that the appearance anchor is always structurally close to the target pose. This dynamic matching drastically reduces the generative burden on the diffusion model. Consequently, as shown in the bottom row of Fig.~\ref{fig:fig9}, our method yields highly faithful pose alignment and strictly physically plausible human structures, completely avoiding the severe anatomical distortions observed in the baseline.

\subsection{Case Study: Dynamic Reference Retrieval in Action}

To vividly illustrate \emph{which} frames our SCM module selects and \emph{why} this dynamic selection is highly effective, we provide a concrete case study in Fig.~\ref{fig:fig10}. 

When relying solely on the initial reference image, diffusion baselines often struggle with ambiguous or partially occluded garment topologies. In this example, the character is wearing a long-sleeved top. However, as the character performs complex arm movements, the initial static reference fails to provide sufficient explicit guidance for the arm regions. Consequently, the baseline (DynamiCtrl) suffers from severe topological instability. As highlighted by the red bounding boxes in the second row of Fig.~\ref{fig:fig10}, the generative model becomes deeply confused, causing the character's arms to violently flicker between bare skin, short sleeves, and long sleeves across adjacent frames.

FrameCache elegantly resolves this ambiguity through its dynamic memory. As the sequence progresses, the SCM module successfully caches high-quality generated frames where the complete long-sleeve topology is explicitly and clearly visible. When an upcoming driving pose involves significant arm movements, the \emph{Match} stage automatically retrieves these optimal cached frames rather than defaulting to the ambiguous initial image. By conditioning the generation on these pose-aligned, topologically explicit references (shown in the third row of Fig.~\ref{fig:fig10}), our method drastically reduces the inferential guesswork of the model. As a result, FrameCache strictly maintains the correct long-sleeve appearance throughout the entire long sequence, completely eliminating the temporal flickering and garment hallucination.

\end{document}